\documentclass[letterpaper, 10 pt, conference]{ieeeconf}  

\usepackage{graphicx}
\usepackage{subfigure}
\usepackage{hyperref}

\usepackage{amsmath}
\usepackage{verbatim}
\usepackage[section]{placeins}
\usepackage{float} 
\usepackage{color}
\usepackage{multirow}
\usepackage{subfigure}
\makeatletter
\newcommand{\printfnsymbol}[1]{%
	\textsuperscript{\@fnsymbol{#1}}%
}

\usepackage{color}

\IEEEoverridecommandlockouts                              

\title{\LARGE \bf
Few-Shot Relation Learning with Attention \\ for EEG-based Motor Imagery Classification}

\author{Sion An$^{1}$, Soopil Kim$^{1}$, Philip Chikontwe$^{1}$ and Sang Hyun Park$^{1}$
\thanks{This work was supported in part by the Daegu Gyeongbuk Institute of Science and Technology (DGIST) Research and Development Program of the
Ministry of Science, and the ICT under Grant 19-RT-01.\newline
$^{1}$Department of Robotics Engineering, DGIST, Daegu, South Korea. \{sion\_an, soopilkim, philipchicco, shpark13135\}@dgist.ac.kr}%
}

\begin{document}

\maketitle
\thispagestyle{empty}
\pagestyle{empty}

\begin{abstract}
Brain-Computer Interfaces (BCI) based on Electroencephalography (EEG) signals, in particular motor imagery (MI) data have received a lot of attention and show the potential towards the design of key technologies both in healthcare and other industries. MI data is generated when a subject imagines movement of limbs and can be used to aid rehabilitation as well as in autonomous driving scenarios. Thus, classification of MI signals is vital for EEG-based BCI systems. Recently, MI EEG classification techniques using deep learning have shown improved performance over conventional techniques. However, due to inter-subject variability, the scarcity of unseen subject data, and low signal-to-noise ratio, extracting robust features and improving accuracy is still challenging. In this context, we propose a novel two-way few shot network that is able to efficiently learn how to learn representative features of unseen subject categories and how to classify them with limited MI EEG data. The pipeline includes an embedding module that learns feature representations from a set of samples, an attention mechanism for key signal feature discovery, and a relation module for final classification based on relation scores between a support set and a query signal. In addition to the unified learning of feature similarity and a few shot classifier, our method leads to emphasize informative features in support data relevant to the query data, which generalizes better on unseen subjects. For evaluation, we used the BCI competition IV 2b dataset and achieved an 9.3\% accuracy improvement in the 20-shot classification task with state-of-the-art performance. Experimental results demonstrate the effectiveness of employing attention and the overall generality of our method.  

\end{abstract}

\graphicspath{{./figures/}}

\section{INTRODUCTION}
Electroencephalography (EEG) is an easy and cheap technique for recording brain activity commonly used for brain-computer interfaces (BCI), enabling communication between the brain and external devices. The high temporal resolution of EEG signals is a key characteristic that makes them useful for research and diagnosis related to brain disorders. Many studies such as robotic arm control~\cite{meng2016noninvasive,ramos2012proprioceptive}, wheelchair control~\cite{ng2014development,galan2008brain}, autonomous driving~\cite{yang2018driving} have been conducted by capturing human brain signals through the EEG. Among several types of EEG, motor imagery (MI) signals have attracted attention and are flexible for discriminating brain activity. MI is produced as a response to thinking tasks \textit{i.e.}, intention to perform hand or leg movements. Accordingly, automated MI classification has been addressed using machine learning~\cite{bashar2016classification,hu2011feature,li2017adaptive,yang2014motor,saa2012latent,ang2008filter,bentlemsan2014random, tiwari2019multiclass} as well as deep learning techniques~\cite{dai2019eeg,zhang2019novel,amin2019multilevel,zhang2019augmentation,wang2018data, dai2020hs, ha2019motor}. However, the low signal-to-noise ratio (SNR), limited spatial resolution and complex dynamics of MI present challenges for the accurate classification and analysis of EEG data. Moreover, robust deep learning models require large amounts of labeled data to be effective given large patient-to-patient signal variations for different behaviors.


To address these limitations, we propose to model EEG-based MI classification as a few shot classification task. Few-shot classification addresses the scenario where a classifier must adapt to accommodate classes which are not seen during training given only a few labeled samples per class (or subject). Thus, we propose a model that learns how to learn representative features of unseen subject categories and be able to accurately classify them using paired samples from support and query data obtained from other subjects. The proposed method consists of feature embedding, attention and relation modules that are all connected in an end-to-end framework. In our framework, relevant features from the support and query data are obtained via the embedding module. Then, class representative features are generated via an attention module that discovers key features related to the query signal among the support data in an episode. Lastly, a relation module predicts the label of a query sample in the embedding space. Our method is able to model the relationship between several paired samples from the support and query sets among different subjects during training; further enabling generalization to classify the query signal of an unseen subject in testing even with few labeled samples.

The main contributions of this work are as follow: (1) In this study, we empirically show the benefit of few-shot learning applied to EEG-based MI classification with improvements in performance. To the best of our knowledge, this work is the first to show the application of few-shot learning in EEG-based MI classification. (2) We introduce a novel few-shot attention technique that can emphasize important signals among the support set to predict the label of a query signal. (3) We integrate an appropriate 1D convolutional neural network (CNN) in the few-shot learning embedding module, which divides a signal into three signals with different frequency ranges and adopts varying kernel sizes. The proposed method can be applied to not only EEG but also various bio-signal $N$-way $K$-shot classification problems.


\section{RELATED WORKS}
\noindent{\textbf{Conventional methods:}} Several methods consisting of feature extraction and classification steps have been proposed for EEG classification. For example, \cite{bashar2016classification} extracted features using short time fourier transform (STFT).~\cite{hu2011feature,li2017adaptive} extracted features representing time-frequency information using wavelet packet decomposition (WPD).~\cite{yang2014motor,saa2012latent,ang2008filter,bentlemsan2014random, tiwari2019multiclass} decomposed the EEG signals into spatial patterns and extracted features using common spatial pattern (CSP). Given the extracted features, classification was performed using machine learning models such as KNN classifier, support vector machine (SVM), XGBoost, or random forest. These methods required heuristic parameter setting such as defining frequency bands, and thus showed limited performances since the optimal parameters were different for each task or subject.

\noindent{\textbf{Deep learning based methods:}} Deep learning based MI EEG classification methods have shown high performance in literature.~\cite{amin2019multilevel} applied 1D CNNs to extract multi-level features and predict the label. In contrast, \cite{dai2019eeg} transformed a time domain signal into frequency domain using STFT, and then applied CNN on the frequency intensity. \cite{zhang2019novel} proposed a network which contains both CNN and long-term short-term memory model (LSTM) to handle sequential time domain data. In a follow-up study, \cite{dai2020hs} proposed a CNN with hybrid convolution scale (HS-CNN) which separates a signal into three signals using bandpass filters with 4$\sim$7Hz, 8$\sim$13Hz, 13$\sim$32Hz frequency bands, and feed them into convolution layers with different filter sizes. The features including different semantic information were concatenated and then MI classification was performed. Further, as an auxiliary, data augmentation~\cite{zhang2019novel,wang2018data,dai2020hs} or attention mechanism~\cite{zhang2019classification,chen2019hierarchical} were also proposed.~\cite{zhang2019classification} applied the attention module to LSTM to utilize long range information for EEG-based hand movement classification.~\cite{chen2019hierarchical} used attention modules to focus on important part of continuous signal as well as to find an important trial among the whole trials for EEG-based emotion classification. 

Despite the improved performance over the conventional methods, the deep learning methods often fail when training samples per subject are limited. Thus, a huge number of training samples need to be obtained from each target subject to train the robust model. To overcome these drawbacks, we propose a relation based few-shot learning method under a single framework that is able to accurately predict the label of query signal using only a few labeled samples. 

\noindent{\textbf{Few shot learning:}} To the best our knowledge, literature using few-shot learning methods for EEG-based MI classification is limited. Although not applied for EEG-based MI classification, the work of \cite{burrello2019hyperdimensional} extracted features from intracranial electroencephalography (iEEG) signals using local binary patterns and learned prototype vectors representing a class in a hyper-dimensional space. Then, they performed classification based on the distance between prototype vectors and a vector from query signal. Although they considered the one-shot learning problem, this method did not provide an end-to end learning framework.~\cite{narwariya2020meta} recently proposed a deep neural network with triplet loss to address the classification of various time series data including ECG recordings. This method demonstrates the possibility that few-shot learning can work effectively for signal analysis problems; however, they only learn feature embedding networks so that embedded features can be categorized by a fixed nearest neighborhood classifier.

In this paper, we propose a novel relation based few-shot learning method that can learn a non-linear comparator by learning end-to-end relationships between the support data distribution and query signal without calculating the distance between embedding vectors. The structure of the proposed network closely resembles the recent network proposed for image classification~\cite{sung2018learning}. However, unlike the method of \cite{sung2018learning} that uses a conventional CNN for feature extraction and simply averages the embedding vectors for relation computation, we integrate a modified HS-CNN that is suitable for 1D signal analysis in the embedding module as well as incorporate an attention module that enables the emphasis of important signals in the supporting set.

\section{METHOD}

\begin{figure*}[t!]
	\centering
	\includegraphics[width=1.0\textwidth]{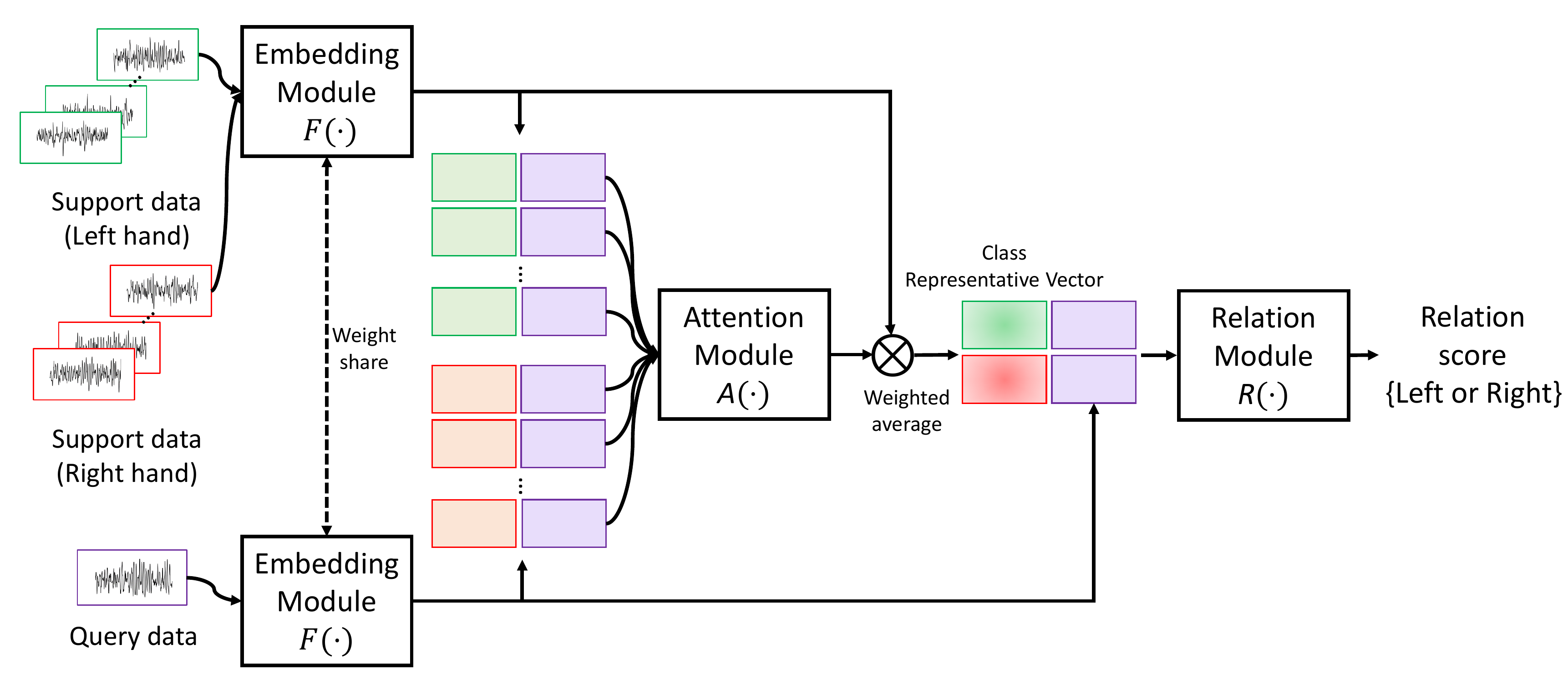}
		\vspace{-0.5cm}
	\caption{Overview of the proposed few-shot relation network. Our framework has three key modules: embedding, attention and a relation module. Given support and query data, we obtain feature vectors concatenated at several stages in our pipeline via an embedding module $F(\cdot)$ which are feed to an attention module $A(\cdot)$ that focuses key features related to the query data based on attention scores. Attended features are further feed to a relation  module $R(\cdot)$ that predicts relation scores from representative vectors and query features to obtain the label of the given query sample. Based on the predictions, we update the weights of the entire framework in a single training episode.}
	\label{fig:network}
\end{figure*}

Our proposed $2$-way $K$-shot learning framework is shown in Fig.~\ref{fig:network}. The framework consists of three modules: an embedding module, an attention module, and a relation module. The embedding module extracts semantic features for classification from input EEG signals. Given the extracted features, attention module predicts an attention score for each support sample using both support and query features. A class-representative vector is calculated for each class using a weighted average of $k$ support features based on the predicted attention scores. Finally, the relation module predicts relation scores given the class-representative vectors and query features. The class with the largest relation score is predicted as the label of the query.
During training, the proposed network is trained end-to-end using pairs of a support set and a query signal from subjects in training data. For testing, we predict the label of a query signal using $k$ labeled support signals from an unseen subject. Though the figure shows an example of $2$-way $K$-shot learning, the proposed method can be extended to $N$-way $K$-shot learning.

\begin{figure}[t] 
\begin{center}
\includegraphics[width=1.0\linewidth]{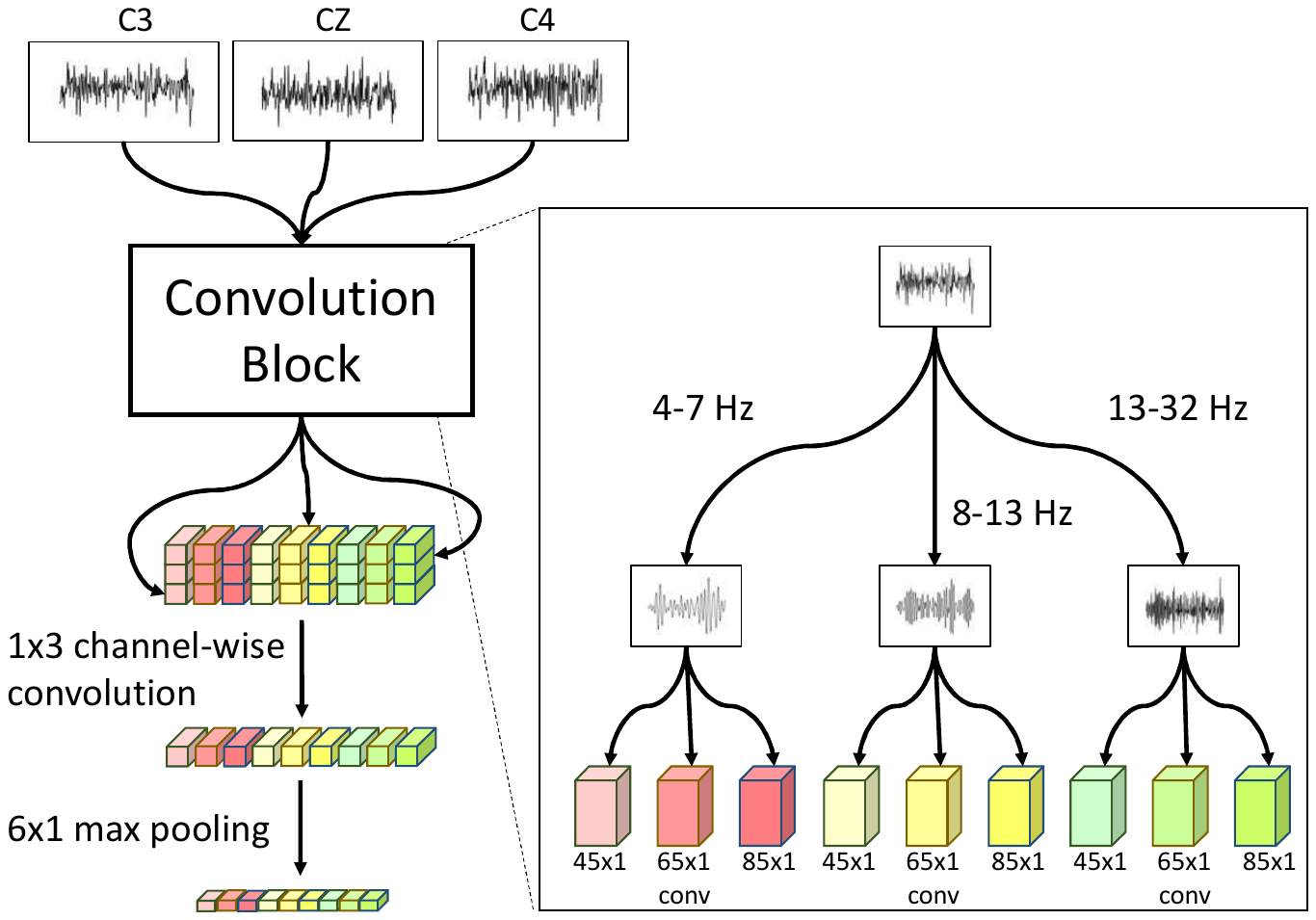}
\end{center}
\vspace{-0.3cm}
\caption{Illustration of the proposed embedding module $F(\cdot)$.}
\vspace{-0.3cm}
\label{fig:embedding}
\end{figure}

\subsection{Embedding Module}

The embedding module $F(\cdot)$ takes a pair of $k$ support and a query data as input and extracts semantic features for classification using convolutional layers. To extract meaningful features from a noisy one-dimensional signal, we modified HS-CNN~\cite{dai2020hs} which separates the data into 3 different frequency bands and extract features per band instead of using convolution layers directly on the data. Specifically, the data $x$ obtained from C3, CZ, and C4 electrodes are divided into 4$\sim$7Hz, 8$\sim$13Hz, 13$\sim$32Hz frequency band signals $x_b$ ($b$ = 4$\sim$7, 8$\sim$13, 13$\sim$32) using a $4^{th}$ order butterworth bandpass filter. For each $x_b$, features were extracted using three convolution layers with different kernel sizes (\textit{i.e.}, 45$\times$1, 65$\times$1, and 85$\times$1). The three features obtained are concatenated and passed through another convolution layer with kernel size of 1$\times$3 to extract higher-level features between the C3, CZ and C4 electrodes. Then, the features are passed through a max pooling layer with a kernel size of 6$\times$1 with a stride 6$\times$1. Finally, the embedding feature $z=F(x)$ is obtained by concatenating all features. Fig.~\ref{fig:embedding} shows the embedding network.


\subsection{Attention Module}
The support set of $N$-way $k$-shot learning contains $N \times k$ data samples. Given extracted features for each data sample from the embedding module, the attention module $A(\cdot)$ computes attention scores to get $N$ class-representative features most related to the query prediction. In the original relation network~\cite{sung2018learning}, class representative features are defined as a summation of extracted features from support data. However, this approach does not effectively utilize the data for query prediction since all support data samples are reflected equally in the class-representative feature. Though the approach is an easy example that may explain why a given query set belongs to a certain class, it may lead to inaccurate prediction if noisy samples exist in the support set. 

To address this, we hypothesize that an attention-based approach may better help our model to focus on key support samples. Thus, we concatenate the features of each support feature $z^S_{i,j}$($i = 1, 2,…, N; j = 1, 2,…, k$) and query feature $z^Q$ in the channel direction and use the feature $z^{SQ}_{i,j}= z^S_{i,j} \oplus z^Q$ as input for the attention module to predict the attention score $a^{S}_{i,j} = A(z^{SQ}_{i,j})$, where $\oplus$ is the concatenation operation. The attention module consists of convolutional layers using 16$\times$1 and 4$\times$1 kernels, a global average pooling layer, and 64 and 1-dimensional fully connected layers to consider both global and local features.

For each class, the attention-weighted feature $\overline{z^{S}_{i}}$ is computed as:
\begin{equation} \label{eq::attention}
\vspace{-0.2cm}
\overline{z^{S}_{i}} = {\sum_{j=1}^k a^{S}_{i,j}*z^S_{i,j} \over \sum_{j=1}^k a^S_{i,j}}.
\end{equation}
Then, $\overline{z^{S}_{i}}$ and $z^Q$ are concatenated to obtain the final class-representative feature vector $z^{SQ}_{i}= \overline{z^{S}_{i}} \oplus z^Q$ for each class. Finally, the class-representative features are fed to the relation module to predict the label.

\subsection{Relation Module}
While several few-shot learning methods predict the label based on a distance metric between class-representative features and a query feature, we estimate the label by learning a relation network to distinguish them. The relation module $R(\cdot)$ predicts a relation score $r^{S}_{i}=R(z^{SQ}_{i})$ from $z^{SQ}_{i}$. Finally, the class with largest predicted relation score is taken as a label. To estimate the label from the 1D feature vector, the relation module uses two convolutional layers with 30$\times$1 and 15$\times$1 kernels, a global average pooing layer, and two fully connected layers with dimensions 256 and 100. 

\subsection{Implementation details}
For training, we optimized the model using cross-entropy loss function  as follows:
\begin{equation} 
Loss = - {1 \over N} * \sum_{i=1}^N y_i*log(r^S_i),
\end{equation}
where $y_i$ is 1 if the class of supporting data and that of query data are the same, otherwise 0. Adam optimizer~\cite{kingma2014adam} was used to optimize the parameters. The batch size was set to 100 and the initial learning rate was $10^{-4}$. On every iteration, learning rate decayed with an exponential decay rate of 0.033\%. The model was saved when validation loss was minimum.

\section{EXPERIMENTAL RESULTS}
\subsection{Dataset}
The BCI competition IV 2b dataset~\cite{leeb2008bci} was used to evaluate the performance of proposed network. The dataset contains raw EEG data of 9 subjects for MI classification with approximately 120 trials per experiment, where each subject imagines’ left- or right-hand movement according to instruction. Five experiments were conducted on each subject with a total of 45 experiments collected from 9 subjects. The collected samples were measured at a sampling frequency of 250 Hz on three electrodes C3, CZ, and C4 in accordance with the protocol of International 10-20 system. Among the 5400 trials (\textit{i.e.}, 120$\times$45), we ignored some rejected trials and then used signals from 3.5s to 7s of the remaining trials in this study. From 3 electrodes, we obtained three 875 values and stacked them to form a 875$\times$3 matrix which was used as input to the model. 

\begin{figure*}[t]
	\centering
	\includegraphics[width=1.0\textwidth]{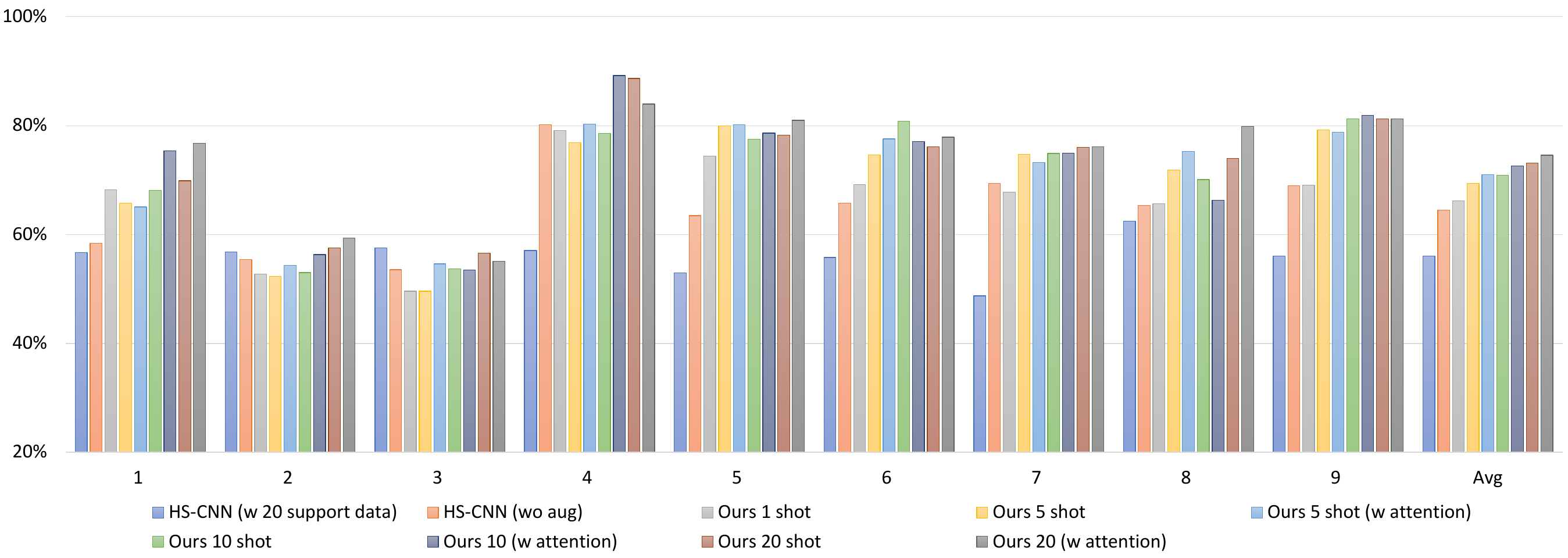}
		\vspace{-0.5cm}
	\caption{Comparison of our few shot learning methods with related methods across 9 validation folds.}
	\label{fig:resultbar}
\end{figure*}

\subsection{Experimental Setting}
For our experiments, 9-fold cross validation was used with the model being trained with 8 subject’s data samples and tested with the remainder per validation episode. Each training set consisted of the first four experiments per subject, with the fifth experiment used as the validation set. The test set consisted of the full set of the remaining subjects. During training, support and query data samples were randomly selected from both the training and validation set at each iteration. For testing, we split whole samples into two groups in each class, 20 samples as support and the rest as query data.

To assess the performance of the proposed model (\textbf{RelationNet-attention}), we evaluated the classification accuracy of $K$ = \{1,5,10,20\} shot classification models. For $K$ = \{1,5,10\} shot experiments, $K$ support samples were randomly selected among the 20 support samples. To assess the impact of the attention, we also evaluated the performance of 5, 10 and 20-shot models without attention (\textbf{RelationNet}). In addition, we compare our method with following two supervised learning models: (1) HS-CNN~\cite{dai2020hs} trained with only 40 samples (20 left and 20 right samples) in the support set of testing subjects (\textbf{HS-CNN-Few}) and (2) HS-CNN trained with all training samples from 8 subjects (\textbf{HS-CNN-All}). Since the results may vary as the support set changes, the same experiment was repeated 10 times and the average was used as the final accuracy for all cases. The accuracy was measured based on the number of true positive and true negative over the entire dataset.

\begin{table}[t] \centering
\caption{Comparison of 9-fold cross validation results of related methods using different $k$ shot settings. Each row shows the average accuracy and standard deviation.} \label{tab:result1}
\begin{tabular}{|c||l|c|c|c|c|c|c|c|c||c|}
 
 \hline
   & Method & Average $\pm$ Std \\ \hline
- & HS-CNN-Few & 56.0 $\pm$ 3.7\% \\ \hline
- & HS-CNN-All & 64.5 $\pm$ 8.2\% \\ \hline
1 shot & RelationNet & 66.2 $\pm$ 9.5\% \\ \hline
5 shot & RelationNet & 69.4 $\pm$ 11.3\% \\ \cline{2-3}
& RelationNet-attention & 71.0 $\pm$ 10.5\% \\ \hline
10 shot & RelationNet & 70.9 $\pm$ 10.9\% \\ \cline{2-3}
& RelationNet-attention & 72.6 $\pm$ 11.7\% \\ \hline
20 shot & RelationNet & 73.1 $\pm$ 10.5\% \\ \cline{2-3}
& RelationNet-attention & 74.6 $\pm$ 10.2\% \\ \hline

\end{tabular}
\end{table}

\subsection{Results and Analysis}
Table 1 shows the overall accuracy with standard deviations of the proposed methods and other methods. Fig. 3 shows the accuracy of the methods for 9 folds as a graph. In all cases, our models achieve better performance than the prior approaches. Specifically, supervised learning-based method reports that classification performance reaches 87.6\% when sufficient training data acquired from the target subject is used \cite{dai2020hs}, but the performance significantly drops as 65.3\% when the model was trained in a cross-subject setting. When the model was trained with a small amount of data (20 samples per class) acquired from the target subject, the accuracy was also limited. On the other hand, the proposed few shot learning technique greatly improved the performance. In the one-shot setting with only one labeled sample per target subject used for training, a 1.7\% improvement in performance was observed compared to HS-CNN-All, 4.9\% in 5 shots, 6.4\% in 10 shots, and 8.6\% in 20 shots. Finally, our approach achieved 74.6\% (+9.3\%) mean accuracy in 20-shot with attention setting. 

The proposed attention module improved the performance consistently. In particular, an average and consistent improvement of 1.6\% can be observed whenever attention is employed in all cases (\textit{i.e.}, the average percent improvement over all the proposed methods when $K$ = \{5,10,20\}). Moreover, by using the attention module, we can infer that the model concentrates on the support set feature representations that are important for query prediction, while at the same time suppress noisy signal features to achieve higher performance. 


We further observed that in the $2^{nd}$ and $3^{rd}$ folds, accuracy scores of both HS-CNN-All and the proposed model were relatively poor (See Fig. 3). These results could be an indication of the nature of EEG data; due to large SNR, it was hard to classify the query data with only a few support sets. We believe significant performance improvements may be observed if noise in the data is sufficiently reduced. For a more effective comparison, the average prediction accuracy, except for 2$^{nd}$ and 3$^{rd}$ folds, was calculated as 67.3\% for HS-CNN-All and 79.5\% for 20-shot RelationNet-attention. In particular, our few shot learning method showed comparable results in all folds except 2nd and 3rd as the std decreased, and also showed a 12.2\% performance improvement over HS-CNN-All.

\subsection{Effect of data augmentation in the few shot setting}
Dai \textit{et al.}~\cite{dai2020hs} report that HS-CNN performance improved when data augmentation was used for training and testing within same subject data. To show the effect of data augmentation, we tested the HS-CNN trained by all training samples with data augmentation. Furthermore, we performed the data augmentation on our few shot learning models (both with and without attetion module) and then confirmed the results. Data augmentation was applied to the training data samples in two stages: (1) time-domain recombination: crop and recombine trials to form new trails, and (2) frequency-domain swapping: band-pass filter trials and generate new trials. The later were repeated multiple times to create increased number of samples (we refer the reader to the \cite{dai2020hs} for more details).

Table 2 shows the mean accuracy and standard deviation. In our study, the improvements gained by applying data augmentation to base method HS-CNN-All were not noteworthy. Furthermore, augmentation was equally ineffective when used in the proposed few shot framework since the quality of augmented samples generated by $K$ images hardly reflected the EEG signals from unseen subjects. As a result, accuracy decreased in most cases when the model was trained with augmented data. On the other hand, interestingly, accuracy increased when the attention module was used even though the accuracy was slightly less than the accuracy of model without data augmentation. Overall, where the base method fails in generality, the proposed method shows consistent improvements across $K$-shot settings and subjects. This result further highlights utility of the proposed attention module in emphasizing important support set features important for query prediction.

\begin{table}[t] \centering
\caption{Comparison of accuracy scores of the proposed and related methods trained with data augmentation.} \label{tab:result2}
\begin{tabular}{|c||l|c|c|c|c|c|c|c|c||c|}
 
\hline
  & Method with augmentation & Average $\pm$ Std \\ \hline
- & HS-CNN-All & 65.3 $\pm$ 9.2\% \\ \hline
5 shot & RelationNet & 68.6 $\pm$ 9.0\% \\ \cline{2-3}
& RelationNet-attention & 69.3 $\pm$ 10.9\% \\ \hline
10 shot & RelationNet & 71.7 $\pm$ 10.6\% \\ \cline{2-3}
& RelationNet-attention & 72.8 $\pm$ 12.5\% \\ \hline
20 shot & RelationNet & 72.2 $\pm$ 10.2\% \\ \cline{2-3}
& RelationNet-attention & 73.6 $\pm$ 12.3\% \\ \hline

\end{tabular}
\end{table}

\subsection{Validation on external dataset}
To assess the generalization performance of our model, we evaluated performance on an independent external dataset. In this experiment, we tested our models on the BCI competition IV 2a dataset~\cite{brunner2008bci}. Note that the models trained with BCI competition IV 2b dataset were directly used for testing without re-training our models. The dataset contains a total of 288 trials with four classes (left hand, right hand, both feet, and tongue movement) for a single experiment. EEG data was measured at a sampling frequency of 250 Hz across a total of 22 electrodes, depending on the protocol of the International 10-20 system. In this study, we used the left- and right-hand trial data of C3, CZ and C4 electrodes in the same location and used 875 values as inputs from 2.5s to 6s of 1163 trials except the rejected trial in the training set. 

The accuracy of the HS-CNN-All was almost 53\%, while the proposed method achieved around 60\%, showing a 6.0\% improvement. Although the results are lower than those reported for subjects of the same dataset (\textit{i.e.}, BCI competition IV 2b dataset~\cite{leeb2008bci}), we could confirm that the proposed few shot methods show comparable generalization even for tasks with slightly different settings without re-training. In addition, the model with attention module also consistently showed better results than the relation network in this experiment.

\begin{table}[t] \centering
\caption{Comparison of accuracy scores of the proposed and related methods on external dataset (BCI $\mathrm{IV}$ 2a~\cite{brunner2008bci}).} \label{tab:result2}
\begin{tabular}{|c||l|c|c|c|c|c|c|c|c||c|}

 \hline
 & Method & Average $\pm$ Std \\ \hline
-& HS-CNN-All & 53.1 $\pm$ 2.0\% \\ \hline
1 shot &RelationNet-attention & 52.9 $\pm$ 6.6\% \\ \hline
5 shot &RelationNet & 56.3 $\pm$ 9.8\% \\ \hline
5 shot &RelationNet-attention & 55.9 $\pm$ 8.3\% \\ \hline
10 shot &RelationNet & 57.5 $\pm$ 9.3\% \\ \hline
10 shot &RelationNet-attention & 58.7 $\pm$ 11.4\% \\ \hline
20 shot &RelationNet & 58.4 $\pm$ 10.4\% \\ \hline
20 shot &RelationNet-attention & 59.1 $\pm$ 11.1\% \\ \hline

\end{tabular}
\end{table}



%
\section{CONCLUSIONS}
In this study, we have proposed a two way few shot classification network which performs well on unseen subjects by applying the concept of few shot learning to the EEG-based MI classification problem. Our model achieves improved performance over baseline methods and gives evidence that few shot learning is able to classify the EEG data of unseen subject even with limited samples. Moreover, we show that using attention can further improves the performance by emphasizing the information of support data that is highly correlated to query data. Empirical results show the general applicability of the proposed method across $K$-shot settings with a novel formulation of relation score as a similarity measure between support and query sets. The proposed model can be easily applied other types of time series data which suffer from large signal variations between subjects. Overall, the simplicity and effectiveness of this approach makes it a promising approach for EEG based MI classification.

\addtolength{\textheight}{-12cm}   









\bibliographystyle{IEEEtran}
\bibliography{EEG_Meta}

\end{document}